\documentclass[
  aps,
  twocolumn,
  pra,
  groupedaddress,
  nofootinbib,
]{revtex4-2}
\usepackage{newtxtext,newtxmath}
\usepackage{microtype}
\usepackage{amsmath}
\usepackage{bm}
\usepackage{physics}
\usepackage{siunitx}
\usepackage{graphicx}
\usepackage{hyperref}
\usepackage[linesnumbered,ruled,vlined]{algorithm2e}
\DontPrintSemicolon
\usepackage{booktabs}
\usepackage{cleveref}
\crefname{equation}{Eq.}{Eqs.}
\crefname{figure}{Fig.}{Figs.}
\crefname{table}{Table}{Tables}
\crefname{algorithm}{Algorithm}{Algorithms}

\newcommand{\br}[1]{\left(#1\right)}
\newcommand{\brs}[1]{\left[#1\right]}
\newcommand{\brc}[1]{\left\{#1\right\}}
\newcommand{\RR}{\mathbb{R}}
\newcommand{\CC}{\mathbb{C}}
\newcommand{\Hcurl}[1]{H(\text{curl};#1)}
\newcommand{\Hcurlt}[1]{H_t(\text{curl};#1)}
\newcommand{\inn}[3]{\langle #1, #2 \rangle_{#3}}
\newcommand{\etal}[1]{#1 \emph{et al.}}
\newcommand{\rnum}[1]{\text{\uppercase\expandafter{\romannumeral #1\relax}}}
\newcommand{\pprime}{{\prime\prime}}

\begin{document}
  \begin{center}
  \mbox{Published in: Phys. Rev. Accel. Beams \textbf{29}, 103802 (2026), doi:10.1103/hjqj-v5yc}
  \end{center}
  
  \title{Numerical quality factor statistics for superconducting rf cavities with spatially inhomogeneous multilayer coatings modeled by Gaussian random fields}

  \author{Aaron Gobeyn}
  \email{aaron.gobeyn@tu-darmstadt.de}
  \affiliation{Institute of Accelerator Science and Electromagnetic Fields (TEMF), TU Darmstadt}

  \author{Wolfgang Ackermann}
  \affiliation{Institute of Accelerator Science and Electromagnetic Fields (TEMF), TU Darmstadt}

  \author{Herbert De Gersem}
  \affiliation{Institute of Accelerator Science and Electromagnetic Fields (TEMF), TU Darmstadt}

  \date{\today}

  \begin{abstract}
    Bulk niobium has long been the material of choice for superconducting radio-frequency applications. An alternative approach
    is the superconductor-insulator-superconductor (SIS) multilayer structure, which enables the use of brittle high-$T_c$ materials,
    such as NbTiN. At present, SIS coatings are limited to flat samples, with the single-cell TESLA cavity representing a key milestone.
    Extending coating processes to non-flat geometries is expected to introduce macroscopic inhomogeneities in coating thickness. We model 
    these variations using Gaussian random fields parametrized by a spatial correlation length and generated by solving a stochastic partial 
    differential equation. The resulting field is incorporated into the boundary condition of the cavity eigenvalue problem, from 
    which quantities of interest---such as the quality factor---are computed. This procedure is repeated for 
    eight correlation lengths, with \num{2048} samples per correlation length, where the resulting quality factors are recorded. Our results 
    show that the quality factors follow a normal distribution. The standard deviation increases with the correlation length and can 
    be statistically distinguished. In contrast, the mean values remain largely unchanged, with only a few significant differences. 
    In extreme cases, depending on the correlation length, the quality factor may differ from the uniform case by \SIrange{2}{6}{\percent}.
  \end{abstract}

  \maketitle

  \section{Introduction}
  
  Bulk niobium (Nb) has long been the material of choice for superconducting radio-frequency (srf) cavities in accelerator 
  physics applications and has therefore been extensively studied to approach its theoretical capabilities. A variety 
  of surface preparation techniques have been developed to improve rf performance. Electropolishing produces smoother 
  surfaces, reducing local enhancement due to roughness, but introduces hydrogen absorption that must be removed 
  by subsequent heat treatment. High-pressure rinsing with ultrapure water is then applied to remove microparticles and 
  reduce field emission \cite{Padamsee2009}.

  More recent advances include nitrogen doping and diffusion \cite{Dhakal2020}, as well as low-temperature baking 
  \cite{Grassellino2018, Wenskat2020}. These methods have significantly improved the quality factor and 
  the achievable accelerating fields. However, such approaches remain fundamentally limited by the critical magnetic field of Nb
  ($H_c{\sim}\SI{200}{\milli\tesla}$). Additionally, the operating temperature of bulk Nb cavities is currently \SI{2}{\kelvin}, 
  and is limited by the critical temperature $T_c$ of Nb. Using higher $T_c$ superconductors could enable operation 
  at \SI{4.2}{\kelvin}, substantially reducing cryogenic costs \cite{ValenteFeliciano2016}.

  Because rf fields penetrate only a very thin surface layer in superconductors 
  (typically less than \SI{1}{\micro\metre} \cite{ValenteFeliciano2016}), the superconducting properties 
  relevant for cavity performance are governed almost entirely by this thin layer. This has motivated the 
  development of thin-film cavities, in which a superconducting coating is deposited onto a bulk substrate. This 
  approach has been successfully implemented in Nb/Cu cavities at CERN facilities such as LEP-II and LHC \cite{Calatroni2006}. An 
  extension of this concept is the use of superconductor-insulator-superconductor (SIS) multilayer structures \cite{Gurevich2006}. Here, 
  a bulk superconducting substrate is coated with a thin insulating layer $\mathcal{O}(\SI{10}{\nano\metre})$ followed by a superconducting 
  thin film $\mathcal{O}(\SI{100}{\nano\metre})$ with higher $T_c$ and $H_c$. The superconducting thin film partially screens the rf field,
  reducing the field strengths encountered by the substrate such that it maintains the superconducting state at higher fields than possible 
  by bulk only. The insulator layer prevents vortex penetration and suppresses Josephson coupling effects \cite{Kubo2016}. One promising 
  candidate is the NbTiN-AlN-Nb multilayer \cite{Burton2016,GonzlezDazPalacio2023,Asaduzzaman2025}. Other approaches include 
  superconductor-superconductor bilayers, such as $\text{Nb}_3\text{Sn}$ on Nb produced via e.g. co-sputtering \cite{Schafer2020,Sayeed2023}, 
  as well as $\text{Nb}_3\text{Sn}$ thin films on Cu substrates, expanding on the successes of Nb/Cu cavities \cite{Fonnesu2026}.

  All of these approaches rely on thin-film deposition techniques, e.g. atomic layer deposition, co-sputtering and dc magnetron sputtering, 
  and are typically investigated on flat samples for quadrupole resonator measurements \cite{Tikhonov2022,Fonnesu2026}. 
  Under such conditions, film thickness is generally uniform, globally speaking. However, extending these techniques to complex 
  geometries, e.g. TESLA-type cavities, is expected to produce large-scale thickness inhomogeneities, with spatial distributions 
  that depend on the particular deposition method. The influences of such inhomogeneities on relevant quantities of interest, e.g. 
  quality factor, remains largely unexplored and is difficult to assess experimentally. In this work, we perform a numerical 
  study to quantify the impact of coating thickness variations on the quality factor of a single-cell TESLA cavity. To model these 
  variations, the coating thickness distributions are emulated through Gaussian random fields (GRFs). In essence, GRFs are a 
  continuous extension of the normal distribution. They consist of an infinite collection of random variables, indexed by 
  spatial coordinates, such that any finite subset follows a joint multivariate normal distribution. Furthermore, a GRF is 
  completely defined by its mean function and covariance kernel \cite{Adler2007}. In this study, the chosen covariance kernel 
  is parametrized by two parameters: the spatial correlation length and the standard deviation. Only the former is relevant, 
  since the generated values are possibly unbound and are consequently rescaled to within a desired thickness range. The correlation length 
  dictates the radius within which neighbouring points display similar thickness values, thereby satisfying the expectation 
  of any deposition method that thicknesses are locally similar. By using GRFs without any additional constraints, we remain 
  agnostic of any particular deposition method. Since only the spatial correlation length is relevant, we sample GRFs over a range 
  of correlation lengths and compare them statistically. 

  The generation of GRFs on complex geometries and the inclusion of them into the electromagnetic eigenvalue problem is not 
  readily available in commercial software. We therefore made use of the open-source software 
  FEniCSx \cite{Baratta2023,Scroggs2022a,Scroggs2022b,Alnaes2014}, which is able to assemble the finite element matrices 
  from the weak formulation of the problem. We outline a general simulation scheme in \cref{fig:flowchart} that is 
  applicable to all of the mentioned thin-film approaches, though we apply it specifically to the SIS structure NbTiN-AlN-Nb for
  illustrative purposes.

  \section{Methods}
      
  \subsection{Eigenproblem formulation}

  In order to obtain the quality factor of the cavity, we use the finite element (FE) method. In particular, we numerically 
  solve the eigenvalue problem:
    \begin{align}
      \label{eqn:curlcurl}
      &\curl (\curl \bm{E}) = \br{\frac{\omega}{c}}^2 \bm{E}, &\text{on } \Omega\\
      \label{eqn:leontovich}
      &\bm{n} \times \bm{E} = Z(\bm{r},\omega) \brs{ \bm{n} \times \br{\bm{n} \times \bm{H} } }, &\text{on } \partial \Omega
    \end{align}
  where $\bm{E}$ and $\bm{H}$ denote the electric and magnetic field strengths, respectively, $\omega \in \CC$ is the angular frequency, 
  $c$ is the speed of light, $\bm{n}$ is the normal to the surface $\partial \Omega$ and $Z(\bm{r},\omega)$ is the frequency-dependent and 
  position-dependent surface impedance. The former, \cref{eqn:curlcurl}, is obtained from Maxwell's equations in frequency 
  domain, without sources, and with homogeneous, linear and isotropic constitutive relations. The latter, \cref{eqn:leontovich}, is the Leontovich surface impedance boundary condition (SIBC) 
  \cite{Yuferev2018}. The boundary conditions can be specified fully in terms of the electric field through Faraday's law 
  $\bm{H} = (j/\omega\mu) \curl \bm{E}$. To write down the weak formulation, we define the space
    \begin{equation}
      \Hcurlt{\Omega} = \brc{ \bm{u} \in \Hcurl{\Omega} \mid \gamma_t(\bm{u}) \in L^2(\partial \Omega)^3 }
    \end{equation}
  where the tangential trace operator $\gamma_t$ is defined by
    \begin{equation}
      \gamma_t: \Hcurl{\Omega} \to H^{-1/2}(\text{div}_{\partial\Omega};\partial\Omega): \bm{u} \mapsto \bm{n} \times \bm{u} \mid_{\partial \Omega}
    \end{equation}
  We further define the inner product
    \begin{equation}
      \langle .,\!. \rangle_{\Omega}: L^2(\Omega)^3 \times L^2(\Omega)^3: (\bm{u}, \bm{v}) \mapsto \int_{\Omega} \dd \mu \, \bm{v}^{\dagger} \bm{u}
    \end{equation}
  where $\dd \mu$ is the appropriate Lebesgue measure and $\bm{v}^\dagger$ denotes the Hermitian conjugate. 
  Finally, we assume that, for all $\omega \in \CC$, the function $Z^{-1}(\bm{r},\omega) \in L^{\infty}(\partial\Omega)$, such we may define the bounded operators 
    \begin{equation}
      M_{Z^{-1}}^\omega: L^2(\partial\Omega)^3 \to L^2(\partial\Omega)^3: \bm{u} \mapsto Z^{-1} \bm{u}
    \end{equation}
  whose action is effectively multiplication by $Z^{-1}$. The weak formulation then reads: find
  $\bm{u} \in \Hcurlt{\Omega}$ and $\omega \in \CC$ such that $\forall \bm{v} \in \Hcurlt{\Omega}$ \cite{Monk2003, Brenner2008}:
    \begin{equation}
      \label{eqn:conteigprob}
      \begin{aligned}
        \inn{\curl \bm{u}}{\curl \bm{v}}{\Omega} &- j Z_0 \frac{\omega}{c} \inn{M_{Z^{-1}}^\omega \gamma_t(\bm{u})}{\gamma_t(\bm{v})}{\partial\Omega} \\
        &- \br{\frac{\omega}{c}}^2 \inn{\bm{u}}{\bm{v}}{\Omega} = 0
      \end{aligned}
    \end{equation}
  where $Z_0$ is the impedance of free space. The domain $\Omega$ is discretized into tetrahedral elements, and the $\Hcurlt{\Omega}$ space is approximated via Nédélec
  basis functions of the first kind $\brc{\bm{w}_i}_{i=1,\dotsc,N}$, where $N$ denotes the number of degrees of freedom (DOF) \cite{Nedelec_1980}. 
  The test and trial functions are expanded in this basis, i.e.
    \begin{equation}
      \bm{u}(\bm{r}, \omega) = \sum_{i = 1}^N x_i(\omega) \bm{w}_i(\bm{r})
    \end{equation}
  which reduces the continuous problem \cref{eqn:conteigprob} to the discrete eigenvalue problem
    \begin{equation}
      \label{eqn:disceigprob}
      \brs{\bm{K} - j Z_0 \frac{\omega}{c} \bm{B}(\omega) - \br{\frac{\omega}{c}}^2 \bm{M} } \bm{x} = 0
    \end{equation}
  where the components of the stiffness matrix $\bm{K}$, boundary matrix $\bm{B}(\omega)$ and mass matrix $\bm{M}$ are given 
  by 
    \begin{align}
      K_{ij} &= \inn{\curl \bm{w}_i}{\curl \bm{w}_j}{\Omega}, \\ 
      B_{ij}(\omega) &= \inn{M_{Z^{-1}}^\omega \gamma_t(\bm{w}_i)}{\gamma_t(\bm{w}_j)}{\partial\Omega}, \\ 
      M_{ij} &= \inn{\bm{w}_i}{\bm{w}_j}{\Omega}
    \end{align}
  for $i,j = 1, \dotsc, N$. The resonance frequency of an eigenmode $f$ and the quality factor $Q$, which 
  quantifies the damping, can be obtained from the eigenvalue $\omega$ via
    \begin{equation}
      \label{eqn:f_and_q}
      f = \frac{1}{2 \pi} \Re(\omega), \quad Q = -\frac{\Re(\omega)}{2 \Im(\omega)}
    \end{equation}
  We solve the eigenvalue problem \cref{eqn:disceigprob} via a fixed point iteration as detailed in \cref{alg:fixedpoint}. We choose 
  an initial guess, then repeatedly reduce the non-linear problem \cref{eqn:disceigprob} to a generalized eigenvalue problem. The 
  reduced problem is subsequently solved numerically. The solution of the problem yields the 
  eigenpair for the next iteration. This process is repeated until relative convergence of the eigenvalue with tolerance $\varepsilon$.

  \begin{algorithm}[!htbp]
    Choose an initial guess $\omega_0 \in \CC$ and set $i = 0$ \;
    \Repeat{$|\omega_i - \omega_{i - 1}|/|\omega_i| < \varepsilon$}{%
      Assemble the auxiliary matrix
      \[
        \bm{A}_i = \bm{K} - j Z_0 \frac{\omega_i}{c} \bm{B}(\omega_i) 
      \] \;
      Solve the generalized eigenvalue problem
      \[
        \brs{ \bm{A}_i - \br{\frac{\omega}{c}}^2 \bm{M} } \bm{x} = 0
      \] \;
      Extract the converged eigenpair $(\omega_{i+1}, \bm{x}_{i+1})$ closest to the previous eigenpair $(\omega_i, \bm{x}_i)$. \;
      Increment $i \leftarrow i + 1$ \;
    }

    \caption{Fixed point iteration for solving the non-linear eigenvalue problem for a single eigenmode}
    \label{alg:fixedpoint}
  \end{algorithm}

  \subsection{Surface impedance model}

  \begin{figure}[htbp]
    \includegraphics[width=\linewidth]{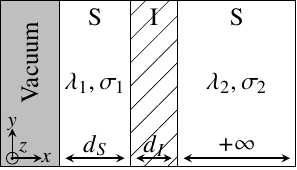}
    \caption{SIS multilayer structure consisting of a superconducting coating layer of thickness $d_S$ with London penetration depth $\lambda_1$ and 
    normal conductance $\sigma_1$, followed by an insulator layer of thickness $d_I$ whose dielectric properties are considered negligible, all 
    layered on top of a superconducting substrate with material parameters $\lambda_2$ and $\sigma_2$ which is considered effectively 
    infinitely thick. The layers are planar and parallel to the $yz$-plane and perpendicular to the $x$-axis.}
    \label{fig:multilayer}
  \end{figure}
  
  Before the eigenvalue problem \cref{eqn:disceigprob} can be solved, a surface impedance model, i.e. description of $Z(\bm{r},\omega)$, must be 
  specified. The material assigned to the boundary $\partial \Omega$ is a SIS multilayer, where 
  the coating thickness is inhomogeneous over the boundary. However, due to the order of magnitude difference in length scales, 
  ${\sim}\SIrange{1}{10}{\milli\metre}$ for the size of mesh cells on the boundary and ${\sim}\SIrange{10}{100}{\nano\metre}$ for the 
  layer thicknesses, we view each boundary cell as a semi-infinite space as illustrated in \cref{fig:multilayer}, with varying thickness 
  parameter $d_S$. Hence we need to determine $Z(\bm{r},\omega)$ for such a semi-infinite space, at a specific boundary point $\bm{r} \in \partial \Omega$. 
  The model by \etal{Kubo} \cite{Kubo2013,Kubo2014,Kubo2016} 
  provides the magnetic fields under reasonable assumptions and simplifications for precisely this setup. The electric and 
  magnetic fields are of the form $\bm{E} = E(x) \exp(-j\omega t) \bm{e}_y$ and $\bm{B} = B(x) \exp(-j\omega t) \bm{e}_z$. Solving Maxwell's 
  equations and the London equations in each layer and imposing continuity in the fields at layer boundaries, one finds that the 
  amplitudes of the magnetic fields in each layer are given by 
  \begin{equation}
    \label{eqn:bfields}
    \begin{aligned}
    &B_\rnum{1}(x) = \gamma B_0 \br{ \cosh(\frac{d_S - x}{\lambda_1}) + r \sinh(\frac{d_S - x}{\lambda_1}) }, \\
    &B_\rnum{2}(x) = \gamma B_0, \\
    &B_\rnum{3}(x) = \gamma B_0 \exp(- \frac{x - d_S - d_I}{\lambda_2})
    \end{aligned}
  \end{equation}
  The magnetic field value $B_0$ is the value at the vacuum and coating layer interface, and the constants $\gamma$ and $r$ are defined by
    \begin{equation}
      \label{eqn:gamma}
      \gamma^{-1} = \cosh(\frac{d_S}{\lambda_1}) + r \sinh(\frac{d_S}{\lambda_1}), \quad r = \frac{\lambda_2 + d_I}{\lambda_1}.
    \end{equation}
  The functions hold in a specific domain, namely $B_\rnum{1}(x)$ is valid for $0 \leq x \leq d_S$, 
  $B_\rnum{2}(x)$ is valid for $d_S \leq x \leq d_S + d_I$ and $B_\rnum{3}(x)$ holds when $x \geq d_S + d_I$.
  We then obtain the surface impedance through the complex Poynting theorem in frequency domain \cite{Jackson1999}
    \begin{equation}
      \begin{aligned}
        \frac{1}{2} \int_\Omega \dd V \, \br{\bm{E} \cdot \bm{J}^\ast} &+ 2j\omega \int_\Omega \dd V\, \br{w_e - w_m} \\
        &+ \oint_{\partial \Omega} \dd S\, \br{\bm{n} \cdot \bm{S}} = 0
      \end{aligned}
    \end{equation}
  where $\Omega$ is an arbitrary volume, $\partial \Omega$ its boundary, $\bm{n}$ the normal vector to that boundary and
    \begin{equation}
      w_e = \frac{1}{4} \bm{E} \cdot \bm{D}^\ast, \quad w_m = \frac{1}{4} \bm{B} \cdot \bm{H}^\ast, \quad \bm{S} = \frac{1}{2} \bm{E} \times \bm{H}^\ast.
    \end{equation}
  Here, $w_e$ and $w_m$ are the energy densities represented by the electric and magnetic components respectively, and $\bm{S}$ is the 
  Poynting vector. We consider the $x \geq 0$ semi-infinite space as the volume $\Omega$. Since the fields must decay to zero 
  at infinity, this means only the $x = 0$ plane contributes to the surface integral. Using the Leontovich boundary condition 
  \cref{eqn:leontovich}, one also finds that
    \begin{equation}
      \bm{n} \cdot \bm{S} \mid_{x = 0} = - \frac{1}{2} Z(\omega) \abs{H_0}^2
    \end{equation}
  Using that the fields depends only on the $x$-direction and rearranging for $Z(\omega)$ one finds that
    \begin{equation}
      \label{eqn:impedance_poynting}
      \begin{aligned}
        Z(\omega) = \frac{1}{\abs{H_0}^2} \bigg[ 
          &\int_{\RR^+} \dd x \, \br{\bm{E}(x)\cdot\bm{J}(x)^\ast} \\
          &+4j\omega \int_{\RR^+} \dd x \, \br{w_e(x) - w_m(x)}
        \bigg]
      \end{aligned}
    \end{equation}  
  The electric field and current density are connected through the modification of Ohm's law by the two-fluid model \cite{London1940}
    \begin{equation}
      \pdv{\bm{J}}{t} = \sigma_n \pdv{\bm{E}}{t} + \frac{1}{\mu_0 \lambda^2} \bm{E}
    \end{equation}
  where $\sigma_n$ is the normal conductivity. In frequency domain this yields,
    \begin{equation}
      \begin{aligned}
        &\bm{J} = \sigma(\omega) \bm{E},\\
        &\sigma(\omega) = \sigma_n + \frac{j}{\mu_0 \omega \lambda^2} := \sigma^\prime + j \sigma^\pprime.
      \end{aligned}
    \end{equation} 
  The quantities appearing in \cref{eqn:impedance_poynting} can be expressed solely in terms of the magnetic fields by using the 
  constitutive relation $\bm{B} = \mu_0 \bm{H}$ and Ampère's law with negligible displacement current:
    \begin{align}
      \label{eqn:e_dot_j}
      \bm{E} \cdot \bm{J}^\ast &= \frac{1}{\mu_0^2 \sigma} \abs{\dv{B}{x}}^2, \\
      \label{eqn:we}
      w_e &= \frac{\varepsilon_0}{4 \mu_0^2 \abs{\sigma}^2} \abs{\dv{B}{x}}^2, \\
      \label{eqn:wm}
      w_m &= \frac{1}{4 \mu_0} \abs{B}^2
    \end{align}
  Substituting \cref{eqn:e_dot_j,eqn:we,eqn:wm} into \cref{eqn:impedance_poynting} we obtain an expression for the 
  surface impedance involving only the magnetic field amplitudes \cref{eqn:bfields}
    \begin{equation}
      \label{eqn:impedance_bfields}
      \begin{aligned}
        Z(\omega) = \frac{1}{\abs{B_0}^2} \bigg[
          &\br{\frac{1}{\sigma} + \frac{j \omega \varepsilon_0}{\abs{\sigma}^2}} \int_{\RR^+} \dd x \, \abs{\dv{B}{x}}^2 \\
          &+ j \omega \mu_0 \int_{\RR^+} \dd x \, \abs{B}^2
        \bigg]
      \end{aligned}
    \end{equation}
  For typical values $\sigma_n{\sim}\SI{e7}{\siemens\per\metre}$, $\omega{\sim}\SI{e10}{\per\second}$ and $\lambda{\sim}\SI{e-7}{\metre}$
  one observes that $\sigma^\prime/\sigma^\pprime{\sim}10^{-3}\ll 1$, so we expand $1/\sigma$ and $1/\abs{\sigma}^2$ up to 
  $\mathcal{O}(\sigma^{\prime 2}/\sigma^{\pprime 2})$, which yields
    \begin{equation}
      \frac{1}{\sigma} \approx \frac{\sigma^\prime}{\sigma^{\pprime 2}} - \frac{j}{\sigma^\pprime}, \quad \frac{1}{\abs{\sigma}^2} \approx \frac{1}{\sigma^{\pprime 2}}
    \end{equation}
  and consequently
    \begin{equation}
      \frac{1}{\sigma} + \frac{j \omega \varepsilon_0}{\abs{\sigma}^2} \approx \sigma_n \mu_0^2 \omega^2 \lambda^4 - j \omega \mu_0 \lambda^2 \br{ 1 - \frac{\omega^2 \lambda^2}{c^2} }
    \end{equation}
  Again, we note that $\omega^2 \lambda^2 / c^2{\sim}10^{-10} \ll 1$ and can therefore be neglected. 
  This further simplifies \cref{eqn:impedance_bfields} to 
    \begin{equation}
      \label{eqn:impedance_bfields_approx}
      \begin{aligned}
        Z(\omega) \approx \frac{1}{\abs{B_0}^2} \bigg[
          &\br{\sigma_n \mu_0^2 \omega^2 \lambda^4 - j \omega \mu_0 \lambda^2} \int_{\RR^+} \dd x \, \abs{\dv{B}{x}}^2 \\
          &+ j \omega \mu_0 \int_{\RR^+} \dd x \, \abs{B}^2
        \bigg].
      \end{aligned}
    \end{equation}
  Naturally, \cref{eqn:impedance_bfields_approx} can be split into three components where the integration bounds are separated according 
  to the domains of $B_{\rnum{1}}(x)$, $B_{\rnum{2}}(x)$ and $B_{\rnum{3}}(x)$, which we denote as
    \begin{equation}
      Z(\omega) = Z_{\rnum{1}}(\omega) + Z_{\rnum{2}}(\omega) + Z_{\rnum{3}}(\omega)
    \end{equation}
  One can show that for the semi-infinite superconductor where $B(x) = B_0 \exp(-x/\lambda)$, \cref{eqn:impedance_bfields_approx} yields
    \begin{equation}
      Z_\infty(\omega) = \frac{1}{2} \sigma_n \mu_0^2 \omega^2 \lambda^3
    \end{equation}
  One then finds that 
    \begin{equation}
      \label{eqn:impedance_coating}
      \begin{aligned}
        Z_\rnum{1}(\omega) = &\gamma^2 \bigg[ 
          Z_\infty^{(\rnum{1})} \bigg\{ -(1 - r^2) \frac{d_S}{\lambda_1} + \frac{1}{2} (1 + r^2) \sinh(\frac{2 d_S}{\lambda_1}) \\
          &+ r \br{ \cosh(\frac{2 d_S}{\lambda_1}) - 1} \bigg\}  + j  \omega \mu_0 (1 - r^2) d_S
        \bigg]
      \end{aligned}
    \end{equation}
  and
    \begin{equation}
      \label{eqn:impedance_ins_and_sub}
      Z_\rnum{2}(\omega) = j \omega \gamma^2 \mu_0 d_I, \quad Z_\rnum{3}(\omega) = \gamma^2 Z_\infty^{(\rnum{3})}(\omega)
    \end{equation}
  We note that the resistive component of the surface impedance is identical to the surface resistance found in Ref. \cite{Kubo2016} 
  through Joule dissipation. There, it is also mentioned that the resistance from dielectric losses is negligible compared 
  to other contributions, concretely when $d_I = \SI{100}{\nano\metre}$, the contribution from dielectric losses is less than 
  $\SI{e-5}{\nano\ohm}$, while the other contributions range from \SIrange{10}{1000}{\nano\ohm}. Furthermore, 
  \cref{eqn:impedance_coating,eqn:impedance_ins_and_sub} has not previously been mentioned for this model, but is required 
  for the eigenvalue problem.

  \subsection{Gaussian random field coating}
  
\begin{figure}[htbp]
    \centering
    \includegraphics[width=\linewidth]{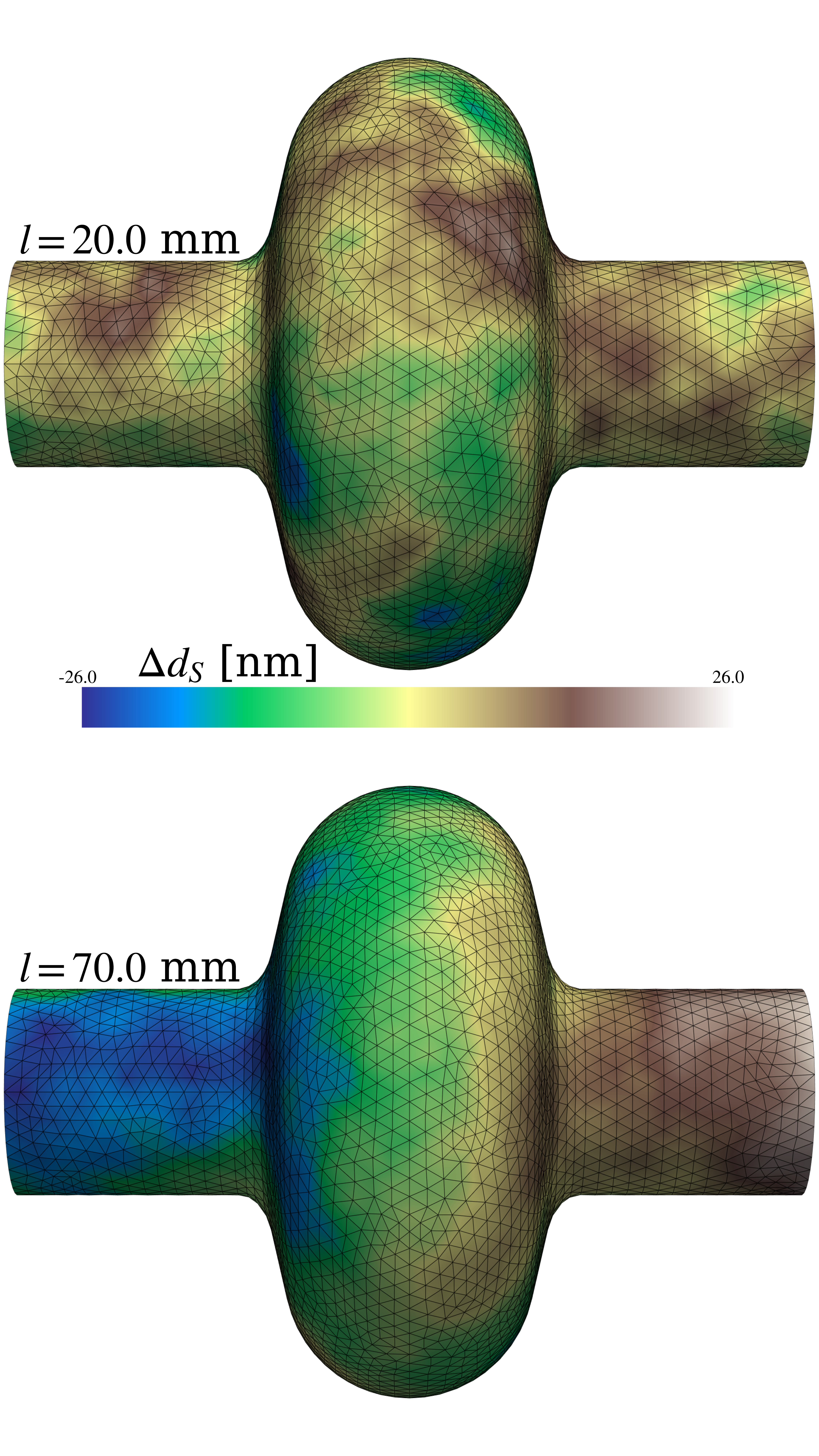}
    \caption{Single cell TESLA cavity with Gaussian random field on the boundary for different spatial correlations lengths. The top plot displays 
    the result for a correlation length of $l = \SI{20}{\milli\metre}$ and the bottom plot for a correlation length of $l=\SI{70}{\milli\metre}$. 
    The field is rescaled to the interval $[-\SI{26}{\nano\metre},\SI{26}{\nano\metre}]$ and represents the deviation from the global 
    mean of the superconducting coating layer $d_S(\bm{r})$.}
    \label{fig:coating}
  \end{figure}

  With the previous computation, we determined the surface impedance $Z(\omega)$ for a particular boundary element. To extend this to a 
  position-dependent surface impedance $Z(\bm{r}, \omega)$, we make the coating thickness position-dependent, i.e. $d_S = d_S(\bm{r})$ for $\bm{r} \in \partial \Omega$.
  For this purpose, we use GRFs. It is known that for zero-mean GRFs with Matérn covariance kernel, a sample can 
  be generated by solving a stochastic partial differential equation (SPDE) \cite{Koh2023}. Here, we will consider the simplest 
  non-trivial Matérn kernel,
    \begin{equation}
      c_h(\bm{r}, \bm{r}^\prime) = s^2 \exp(- \frac{\norm{\bm{r} - \bm{r}^\prime}}{l}),
    \end{equation}
  where $s$ is the standard deviation and $l$ the spatial correlation length. A sample $h(\bm{r}){\sim}\mathcal{G}\mathcal{P}(0,c_h(\bm{r}, \bm{r}^\prime))$ 
  can then be generated by solving the SPDE
    \begin{equation}
      \br{ \frac{1}{l^2} - \laplacian } h(\bm{r}) = \sqrt{\frac{8 \pi s^2}{l}} g(\bm{r}) := \alpha g(\bm{r}) \quad \text{on } \partial \Omega
    \end{equation}
  where $g(\bm{r}){\sim}\mathcal{G}\mathcal{P}(0,\delta(\bm{r} -\bm{r}^\prime))$ is a Gaussian white noise field and $\laplacian$ is 
  understood as the Laplace-Beltrami operator. Using the FE method, we can solve for $h(\bm{r})$ on complex geometries numerically. As 
  before, we define the inner product
    \begin{equation}
      (.,\!.)_{\Omega}: L^2(\Omega) \times L^2(\Omega): (u,v) \mapsto \int_\Omega \dd \mu \, u \bar{v}.
    \end{equation}
  Then the weak formulation reads: for $g(\bm{r}){\sim}\mathcal{G}\mathcal{P}(0, \delta(\bm{r} - \bm{r}^\prime))$, 
  find $u \in H^1(\partial \Omega)$ such that $\forall v \in H^1(\partial \Omega)$
    \begin{equation}
      \frac{1}{l^2} (u,v)_{\partial\Omega} + (\grad u, \grad v)_{\partial\Omega} - \alpha (g,v)_{\partial\Omega} = 0 
    \end{equation}
  Note that no boundary condition was required. Since $\Omega \subseteq \RR^3$ is a Lipschitz domain, $\partial \Omega$ is closed and 
  $\partial(\partial\Omega) = \emptyset$; hence integrals over $\partial(\partial\Omega)$ appearing due to Stokes' theorem vanish without 
  imposing a boundary condition. The space $H^1(\partial\Omega)$ is approximated via Lagrange basis functions $\brc{ \phi_i(\bm{r})}_{i=1,\dotsc,N_s}$, 
  so the test and trial functions are expanded as
    \begin{equation}
      u(\bm{r}) = \sum_{i = 1}^{N_s} u_i \phi_i(\bm{r})
    \end{equation}
  where $N_s$ denotes the scalar DOF. This leads to the discrete linear system
    \begin{equation}
      \label{eqn:gaussian_discrete}
      \br{ \frac{1}{l^2} \bm{M}_S + \bm{K}_S } \bm{u} = \alpha \bm{g},
    \end{equation}
  where the components of the scalar mass matrix $\bm{M}_S$ and scalar stiffness matrix $\bm{K}_S$ are given by:
    \begin{align}
      M_{S,ij} &= (\phi_i,\phi_j)_{\partial\Omega}, \\
      K_{S,ij} &= (\grad \phi_i, \grad \phi_j)_{\partial\Omega}
    \end{align}
  and the vectors $\bm{u}$ and $\bm{g}$ are the expansion coefficients in the FE basis. One can show that 
  $\bm{g}{\sim}\mathcal{N}(0,M_S)$, i.e. $\bm{g}$ is sampled from a zero-mean multivariate normal distribution with the scalar mass matrix as 
  covariance matrix. The position-dependent coating thickness is subsequently defined by
    \begin{equation}
      d_S(\bm{r}) = \hat{d}_S + \Delta d_S(\bm{r})
    \end{equation}
  where $\hat{d}_S$ is a global average coating thickness, and $\Delta d_S(\bm{r})$ is a rescaling of a sample
  $h(\bm{r}){\sim}\mathcal{G}\mathcal{P}(0, c_h(\bm{r}, \bm{r}^\prime))$ to a desired interval. Note that the rescaling 
  effectively makes the standard deviation $s$ irrelevant. We illustrate $\Delta d_S(\bm{r})$ for different correlation lengths 
  $l$ in \cref{fig:coating}. We observe that, as the name suggests, $l$ controls the length scale over which deviations occur.

  \subsection{Simulation scheme}

\begin{figure}[htbp] 
    \centering
    \includegraphics[width=\linewidth]{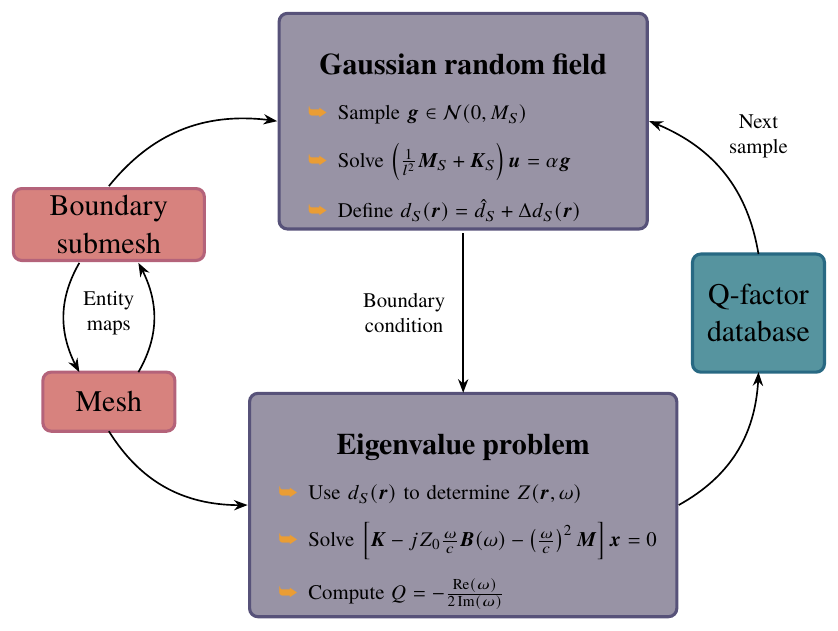}
    \caption{Diagrammatic representation of the simulation scheme. We solve two FE problems, one on the boundary and 
    one on the entire domain. The former yields the boundary condition for the latter. The latter yields the quality factor 
    for a single sample of the coating distribution. Samples are collected in a database and used for statistical analysis.}
    \label{fig:flowchart}
  \end{figure}

  The previously discussed topics are combined into a simulation scheme, which is illustrated as a flowchart 
  in \cref{fig:flowchart}. The domain $\Omega$ is tessellated into a tetrahedral mesh, from which the boundary triangular 
  mesh is extracted, forming the tessellation of $\partial \Omega$. For meshing the TESLA cavity the software Gmsh was used \cite{Geuzaine2009}.
  On the boundary, we then solve \cref{eqn:gaussian_discrete}, which yields a sample of the coating thickness distribution $d_S(\bm{r})$, 
  parameterized by the spatial correlation length $l$. This in turn defines $Z(\bm{r}, \omega)$, which is present in the boundary condition \cref{eqn:leontovich} and hence a component 
  of the eigenvalue problem \cref{eqn:disceigprob}. Solving this problem yields an eigenvalue $\omega$ from which the quality 
  factor $Q$ can be extracted through \cref{eqn:f_and_q}. This process, which involves two FE simulations, returns a single 
  sample of the quality factor. The assembly of the FE matrices in \cref{eqn:disceigprob} and \cref{eqn:gaussian_discrete} was handled 
  by the FE software FEniCSx \cite{Baratta2023,Scroggs2022a,Scroggs2022b,Alnaes2014}. Solving the linear system 
  \cref{eqn:gaussian_discrete} was done using PETSc \cite{Lisandro2011}, and solving the eigenvalue problem of 
  \cref{eqn:disceigprob} was done using SLEPc \cite{Hernandez2005}.

  \section{Results}

  \subsection{Setup}
  
  The single cell TESLA cavity is the first complex geometry many of the research groups working on coating srf cavities are 
  targeting \cite{ValenteFeliciano2022}. Therefore, this is the geometry we focus on. The exact parameters used to describe the geometry 
  can be found in the top section of \cref{tab:parameters}, see Ref. \cite{Wanzenberg2001} for how the 
  parameters come together to form the geometry. We further choose to focus on the operating mode, which for the chosen 
  parameters occurs at a resonance frequency of \SI{1.288}{\giga\hertz}.

  For the multilayer material on the boundary, we choose NbTiN-AlN-Nb, i.e. a bulk Nb substrate with an AlN insulator layer and 
  a NbTiN coating. The relevant material parameters thereof can be found in the second and third sections of \cref{tab:parameters}. 
  Note that a number of parameters, namely $\lambda_{\text{NbTiN}}$, $\lambda_{\text{Nb}}$ and $\sigma_{\text{Nb}}$ need to be 
  rescaled to their appropriate values at \SI{4.2}{\kelvin}. For the two-fluid model, the London penetration depth $\lambda(T)$ is 
  scaled from $\lambda(0)$ by \cite{Gorter1934}
    \begin{equation}
      \lambda(T) = \lambda(0) \brs{1 - \br{\frac{T}{T_c}}^4}^{-1/2}
    \end{equation}
  For the normal conductance, we use the residual resistivity ratio. By definition
    \begin{equation}
      \text{RRR} = \frac{\rho(\SI{300}{\kelvin})}{\rho(\SI{4.2}{\kelvin})} = \frac{\sigma(\SI{4.2}{\kelvin})}{\sigma(\SI{300}{\kelvin})}
    \end{equation}
  Hence $\sigma_{\text{Nb}}(\SI{4.2}{\kelvin}) = \text{RRR}_{\text{Nb}} \sigma_{\text{Nb}}(\SI{300}{\kelvin})$.

\begin{figure}[htbp]
    \centering 
    \includegraphics[width=\linewidth]{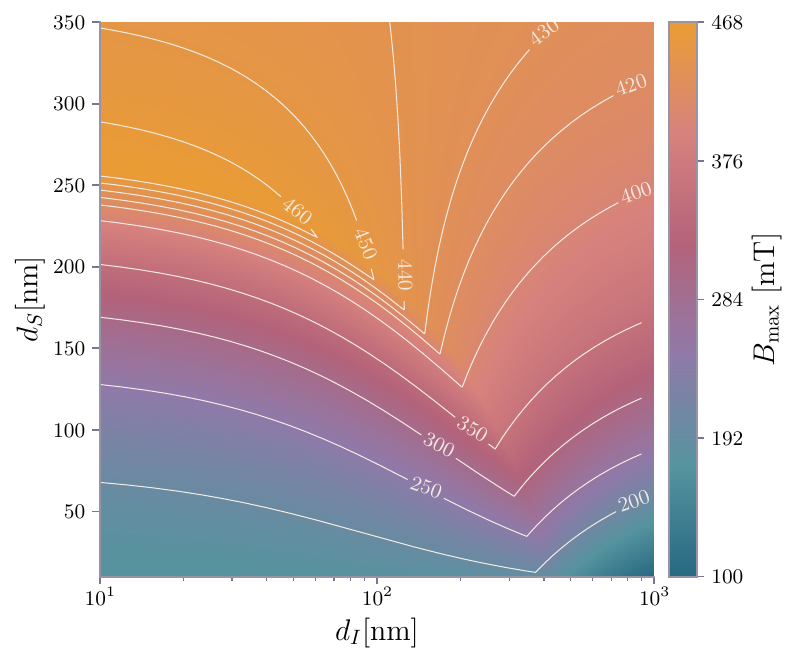}
    \caption{Contour plot of the maximum applicable field $B_{\max}$ depending on thickness of the insulator (AlN) and superconducting 
    coating (NbTiN) layers. The optimum is reached when $d_I = \SI{10}{\nano\metre}$ and $d_S = \SI{260}{\nano\metre}$.}
    \label{fig:contour}
  \end{figure}

  Lastly, the thickness of the insulator and coating layer must be specified. We choose them to optimize for the maximum applicable 
  magnetic field. This corresponds to the highest value of $B_0$ such that the SIS structures maintains the superconducting state. This 
  value depends on the layer thicknesses and is given by \cite{Kubo2016}
    \begin{equation}
      B_{\max} = \min \brc{ \tilde{\gamma}^{-1} B_{\text{sh},\text{NbTiN}}, \gamma^{-1} B_{\max,\text{Nb}} },
    \end{equation}
  where $\gamma$ is defined by \cref{eqn:gamma} and $\tilde{\gamma}$ is given by
    \begin{equation}
      \tilde{\gamma} = \gamma \brs{ \sinh(\frac{d_S}{\lambda_1}) + r \cosh(\frac{d_S}{\lambda_1}) }
    \end{equation}
  Here, $B_{\text{sh},\text{NbTiN}}$ is the superheating field of NbTiN, and $B_{\max, \text{Nb}}$ is the empirical field 
  limit of the bulk Nb substrate. The values of both are listed in \cref{tab:parameters}. We illustrate a contour plot of 
  $B_{\max}$ in \cref{fig:contour}. We find that the optimum is reached when $d_I = \SI{10}{\nano\metre}$ and 
  $d_S = \SI{260}{\nano\metre}$; hence these parameters will be used. The 
  optimum for the coating layer thickness is used for the global average $\hat{d}_S$. For the size of the deviations we take 
  \SI{10}{\percent} of this value, yielding $\Delta d_{S,\max} = \SI{26}{\nano\metre}$. Hence, $d_S(\bm{r})$ lies in the interval 
  $[\num{234},\num{286}] \unit{\nano\metre}$.

  The GRFs used to emulate the coating distribution $d_S(\bm{r})$ depends on the spatial correlation length $l$. We therefore 
  repeat the simulation scheme \cref{fig:flowchart} for a number of correlation lengths, in particular we consider 
  $\SIlist{4;8;12;16;20;24;28;32}{\percent}$ of the cavity length (\SI{265}{\milli\metre}). For each of the correlation length, we 
  sample \num{2048} times in the simulation scheme, in order to obtain a good approximation of the quality factor 
  distributions. The maximum size of a triangle on the boundary of the TESLA cavity mesh was \SI{8.7}{\milli\metre}. The mesh was 
  refined purposefully such that this value is smaller than the smallest correlation length considered (\SI{10.6}{\milli\metre}).

  \begin{table*}[htbp]
    \caption{Parameters used to perform the simulations} 
    \label{tab:parameters}
    \centering
    \begin{tabular}{lcclr}
      \toprule\toprule
      \textbf{Parameter} & \textbf{Value} & \textbf{Unit} & \textbf{Description} & \textbf{Source} \\
      \midrule
      $a$ & \num{35.0} & \unit{\milli\metre} & Iris radius & Ref. \cite{Wanzenberg2001} \\
      $b$ & \num{103.3} & \unit{\milli\metre} & Equator radius & Ref. \cite{Wanzenberg2001} \\
      $h$ & \num{57.7} & \unit{\milli\metre} & Half cell length & Ref. \cite{Wanzenberg2001} \\
      $r_e$ & \num{42.0} & \unit{\milli\metre} & Radius of equator & Ref. \cite{Wanzenberg2001} \\ 
      $r_{iz}$ & \num{12.0} & \unit{\milli\metre} & Horizontal ellipse radius of iris & Ref. \cite{Wanzenberg2001} \\ 
      $r_{ez}$ & \num{19.0} & \unit{\milli\metre} & Vertical ellipse radius of iris & Ref. \cite{Wanzenberg2001} \\
      $l_p$ & \num{75.0} & \unit{\milli\metre} & Length of pipes at either end of the TESLA cavity & ... \\
      \midrule
      $\sigma_{\text{NbTiN}}$ & \num{2.86e6} & \unit{\siemens\per\metre} & Normal conductance of NbTiN at \SI{4.2}{\kelvin} & Refs. \cite{ValenteFeliciano2016,DiLeo1990} \\
      $\lambda_{\text{NbTiN}}$ & \num{180.57} & \unit{\nano\metre} & London penetration depth of NbTiN at \SI{0}{\kelvin} & Ref. \cite{Asaduzzaman2025} \\
      $T_{c,\text{NbTiN}}$ & \num{15.4} & \unit{\kelvin} & Critical temperature of thin layer NbTiN & Refs. \cite{GonzlezDazPalacio2023,Asaduzzaman2025}\\ 
      $B_{\text{sh},\text{NbTiN}}$ & \num{439} & \unit{\milli\tesla} & Superheating field of NbTiN & Ref. \cite{Junginger2018} \\
      \midrule
      $\sigma_{\text{Nb}}$ & \num{6.58e6} & \unit{\siemens\per\metre} & Normal conductance of Nb at \SI{300}{\kelvin} & Ref. \cite{Rumble2017} \\ 
      $\text{RRR}_{\text{Nb}}$ & \num{300} & ... & Residual resistivity ratio of Nb between \SI{300}{\kelvin} and \SI{4.2}{\kelvin}  & Ref. \cite{XFEL2007} \\ 
      $\lambda_{\text{Nb}}$ & \num{39.0} & \unit{\nano\metre} & London penetration depth of Nb at \SI{0}{\kelvin} & Ref. \cite{Maxfield1965} \\
      $T_{c,\text{Nb}}$ & \num{9.23} & \unit{\kelvin} & Critical temperature of bulk Nb & Refs. \cite{ValenteFeliciano2016,Rumble2017} \\
      $B_{\text{max},\text{Nb}}$ & \num{170} & \unit{\milli\tesla} & Maximum applicable field to bulk Nb & Ref. \cite{Kubo2016} \\ 
      \midrule
      $\hat{d}_S$ & \num{260} & \unit{\nano\metre} & Global average thickness of superconducting coating & \cref{fig:contour} \\
      $\Delta d_{S,\text{max}}$ & \num{26} & \unit{\nano\metre} & Maximum deviation of $d_S$ from the mean $\hat{d}_S$ & ... \\
      $d_I$ & \num{10} & \unit{\nano\metre} & Thickness of insulator layer & \cref{fig:contour} \\
      \bottomrule
    \end{tabular}
  \end{table*}

  \subsection{Analysis}
  
  The distribution of quality factors does not necessarily have to be normal. Indeed, if one selects the thickness parameters to minimize 
  the surface resistance, the highest quality factor would be reached. Deviating from this optimum would necessarily maintain or 
  decrease the quality factor, yielding a very skewed distribution. By observing Q-Q plots of the data, we observed that, in this case, 
  the quality factors for each $l$ were normally distributed, i.e. $Q\sim\mathcal{N}(\mu_Q, \sigma_Q)$.

\begin{figure}[htbp]
    \centering
    \includegraphics[width=\linewidth]{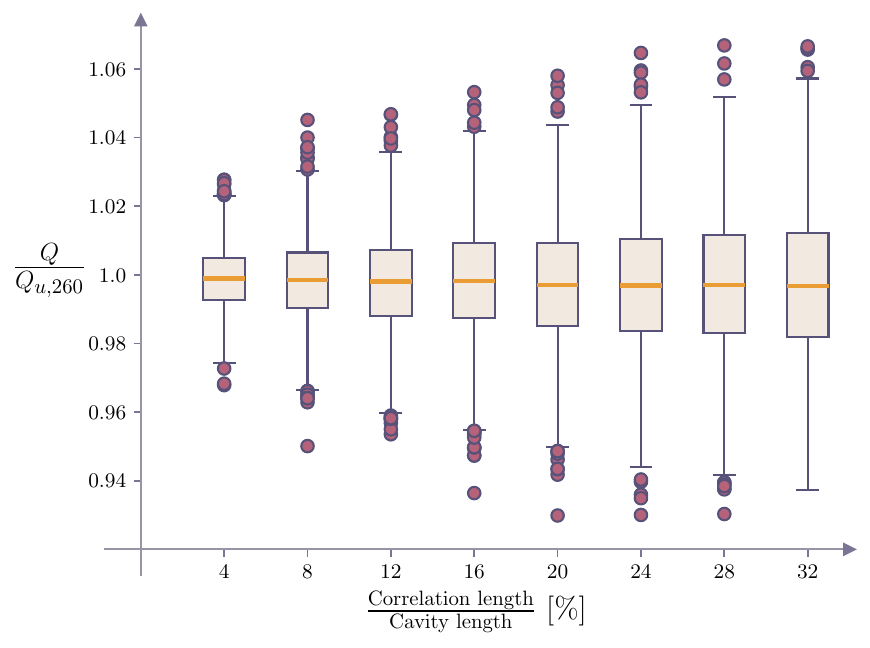}
    \caption{Boxplot of the quality factors for each spatial correlation length. The quality factors are normalized to $Q_{u,260} = \num{1.869647e8}$, the 
    quality factor of a uniform coating with coating thickness $d_S = \SI{260}{\nano\metre}$. As the correlation length increases, 
    the spread becomes larger. In extreme cases, deviations from uniform can reach ${\sim}\SI{6}{\percent}$.}
    \label{fig:boxplot}
  \end{figure}

\begin{figure}[htbp]
    \centering
    \includegraphics[width=\linewidth]{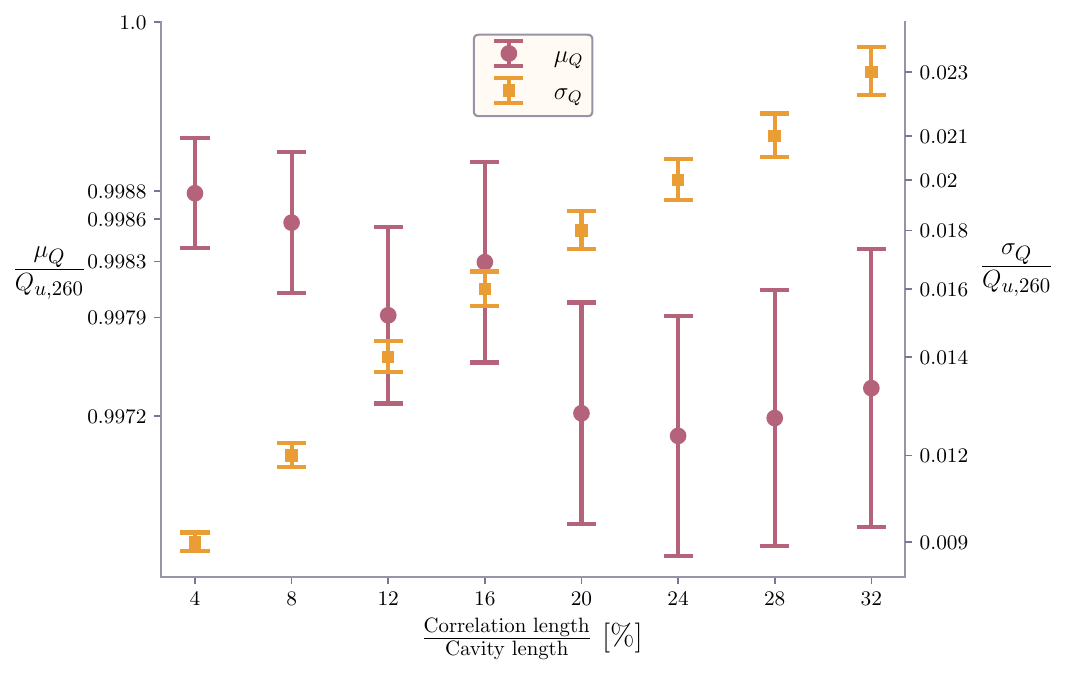}
    \caption{Mean and standard deviation of the quality factor normal distribution for each spatial correlation length. The error bars refer 
    to the \SI{95}{\percent} confidence interval of the quantity. The quantities are normalized to $Q_{u,260} = \num{1.869647e8}$, the quality factor 
    of a uniform coating with coating thickness $d_S = \SI{260}{\nano\metre}$. For all correlation lengths the means lie 
    below the uniform value with respect to the error bars. The error bars of the means have significant overlap, while the error bars
    of the standard deviations have barely any overlap.}
    \label{fig:result}
  \end{figure}

  In \cref{fig:boxplot}, the spread of the quality factors for each $l$ is illustrated using boxplots. We observe 
  that the spread increases with increasing $l$, this trend is corroborated by the $\sigma_Q$ observations in \cref{fig:result}.
  In the most extreme cases, the quality factor deviates from that of a uniformly coated cavity with coating thickness \SI{260}{\nano\metre} by 
  approximately \SI{6}{\percent}. Furthermore, the error bars of the $\sigma_Q$ in \cref{fig:result} show barely any overlap, 
  whereas the error bars of the $\mu_Q$ overlap substantially, particularly for the larger correlation lengths. We additionally 
  observe that the $\mu_Q$ of inhomogeneously coated cavities are consistently smaller than the uniform value, including the error 
  bars. Note that both \cref{fig:boxplot} and \cref{fig:result} are normalized with respect to $Q_{u,260} = \num{1.869647e8}$, 
  which corresponds to the quality factor of a uniform coating with a coating thickness of $d_S = \SI{260}{\nano\metre}$. Applying 
  the classical formula $Q=G/R_s$---using a geometric factor $G=\SI{270}{\ohm}$ \cite{Aune2000} and surface resistance 
  $R_s=\SI{1476}{\ohm}$ calculated via \cref{eqn:impedance_coating,eqn:impedance_ins_and_sub}---yields $Q=\num{1.829e8}$, in 
  good agreement with the simulation. In general, the classical formula is not applicable here since $R_s$ is nonuniform. A similar 
  limitation of the classical formula has been noted in other contexts, such as studies investigating the Q-slope problem 
  \cite{Longuevergne2018,Delayen2018}.

  To verify these observations, hypothesis tests were conducted. First, Levene's test was applied to assess the equality of 
  variances across all eight populations simultaneously \cite{Brown1974}. The resulting $p$-value was extremely small, 
  allowing rejection of the null hypothesis. Thus, the population variances are not all equal. To determine which variances differ, pairwise Levene's tests were 
  performed. The significance level $\alpha = 0.05$ was adjusted for multiple comparisons using the Holm-Bonferroni correction \cite{Holm1979}. 
  All resulting $p$-values were below their adjusted significance levels. The pair of correlation lengths with $p$-value closest to the 
  adjusted significance level was $(\SI{24}{\percent}, \SI{28}{\percent})$, with $p = \num{7.3e-3}$. In 
  \cref{fig:result}, the standard deviations corresponding to these correlation lengths exhibit error bars with slight overlap. From  
  the pairwise comparisons we conclude that all null hypotheses can be rejected. Hence, the variances of all populations are 
  statistically distinct. 

  To compare the means, Welch's ANOVA was used as we are dealing with unequal variances \cite{Welch1951}. We find that the null 
  hypothesis of equal means can be rejected with $p = \num{4.8e-5}$, indicating that not all means are equal. This analysis was followed 
  by a \emph{post hoc} Games-Howell test to examine pairwise equality of the means \cite{Games1976}. We find that the null hypothesis cannot be 
  rejected in nearly all cases, and the few cases for which rejection is possible are only weakly significant. These pairs are: 
  $(\SI{4}{\percent}, \SI{20}{\percent})$, $(\SI{4}{\percent}, \SI{24}{\percent})$ and $(\SI{4}{\percent}, \SI{28}{\percent})$, 
  with corresponding $p$-values \num{0.012}, \num{0.008} and \num{0.033} respectively. Pairs that visually appear distinct 
  in \cref{fig:result}, such as $(\SI{8}{\percent}, \SI{20}{\percent})$, $(\SI{8}{\percent}, \SI{24}{\percent})$ and 
  $(\SI{8}{\percent}, \SI{28}{\percent})$ yield $p$-values close to the 
  significance threshold, namely \num{0.085}, \num{0.054} and \num{0.147} respectively.
  
  \section{Discussion}
  
  From our analysis, we find that the $\mu_Q$ cannot be statistically distinguished in most cases, and only weakly so in a few instances. 
  In contrast, all variances differ significantly in the statistical sense. Hence, increasing $l$ primarily results 
  an increase in $\sigma_Q$, while no significant change in $\mu_Q$ is observed. Nevertheless, we find that the quality factor 
  corresponding to a uniform coating lies significantly above the confidence intervals of all $\mu_Q$ w.r.t the \SI{95}{\percent} confidence interval. 
  Depending on the correlation length, extreme observations show a \SIrange{2}{6}{\percent} increase or decrease in the quality factor. 

  The simulation scheme presented in \cref{fig:flowchart} can be modified and expanded upon in a number of ways; however, the general 
  structure remains the same. First, we used GRFs to emulate the distribution from a coating process. However, 
  depending on the deposition method, certain random fields may not be representative. For example, using 
  dc magnetron sputtering \cite{Calatroni2006, Fonnesu2026}, the cavity is rotated around the sputtering device.
  Consequently, one expects an almost rotationally symmetric thickness distribution rather than an arbitrary GRF sample. 
  More broadly, different deposition techniques, e.g. co-sputtering \cite{Schafer2020,Sayeed2023}, 
  dc magnetron sputtering \cite{Calatroni2006,Fonnesu2026}, high power impulse magnetron sputtering \cite{Arzeo2022}, 
  atomic layer deposition \cite{GonzlezDazPalacio2023}, etc. likely yield distinct representative random fields. Hence, 
  caution is required when applying GRFs without additional constraints. Note, however, that the use of GRFs in 
  \cref{fig:flowchart} can be substituted with any alternative random field generator, such as a surrogate model 
  or a Karhunen-Loève expansion \cite{Ghanem1991}. While this work applies random fields specifically to the thickness of 
  the superconducting thin films, the approach generalizes to any scalar quantity defined on the boundary.

  Second, the random fields enter the eigenvalue problem \cref{eqn:conteigprob} via the surface impedance. While we derived a 
  surface impedance formulation from the work and approximations of Ref. \cite{Kubo2016}, any alternative surface 
  impedance model $Z(\bm{r},\omega)$ can be integrated into the scheme. For example, one could replace the two-fluid model with 
  more advanced descriptions of the optical conductivity of a superconductor, i.e. the complex-valued conductivity 
  \cite{Gurevich2017, Ueki2025}, cf. the models of Mattis-Bardeen \cite{Mattis1958}, 
  Zimmermann \cite{Zimmermann1991}, Eliashberg \cite{Eliashberg1960,Marsiglio2020} or Herman-Hlubina \cite{Herman2021}. 
  Furthermore, this work focuses on macroscale correlation lengths (on the order of millimetres or centimetres) arising from 
  the challenges of depositing onto complex, nonflat geometries. We did not consider microscale correlation lengths 
  (on the order of micrometres), i.e. surface roughness, see e.g. the atomic force microscopy images in Ref. 
  \cite{GonzlezDazPalacio2023}. Resolving these micro-scale geometric features directly through mesh 
  refinement would cause a large increase in computational cost. Instead, surface roughness contributions could 
  be incorporated directly into the surface impedance model, for instance via a correction factor 
  \cite{Wu1994}, or higher-order SIBCs \cite{Senior1960,Yuferev2018}. Modifying the surface impedance model 
  also makes it possible to consider thin-layer approaches other than SIS structures, most notably Nb or Nb$_3$Sn thin films on
  Cu substrates \cite{ValenteFeliciano2016}.

  Finally, while we selected the $Q$ factor as the quantity of interest for statistical analysis, the overall 
  framework can be applied to any quantity derivable from the eigenvalues and electromagnetic eigenmodes. 
 
  \section{Conclusion}
  
  We investigated a single cell TESLA cavity with an inhomogeneously coated NbTiN-AlN-Nb surface. The global average coating thickness 
  was determined according to the configuration that optimizes the maximum applicable field. Coating inhomogeneities were modeled using 
  Gaussian random fields. To analyze their impact, we developed a simulation scheme consisting of two FE simulations: the first generates 
  a Gaussian random field sample on the boundary, while the second incorporates this field into its boundary conditions to compute a quality 
  factor sample. This procedure was carried out for eight spatial correlation lengths, with \num{2048} samples evaluated for each of them. 
  The resulting quality factor distributions were subsequently compared across all correlation lengths. Our analysis revealed that the 
  population means were mostly indifferent, with exceptions being only weakly significant; however, pronounced differences were observed in the 
  standard deviations. We found that the standard deviations increase with correlation length. Furthermore, the quality factor of a 
  homogeneous coating was found to lie above the confidence interval of all population means. Finally, extreme observations indicate that, 
  depending on the correlation length, the quality factor may increase or decrease by approximately $\SIrange{2}{6}{\percent}$.

  \section*{Acknowledgements}

  The authors acknowledge financial support from the Federal Ministry of Research, Technology and Space (BMFTR), Germany, under 
  grant number 05H2024.

  \section*{Data availability}
  The data and software that support the findings of this article are openly available on Zenodo \cite{gobeyn_2026_21474212}.

  \bibliography{main.bib}
\end{document}